\documentclass[aps,twocolumn,groupedaddress,showpacs]{revtex4}
\usepackage[dvips]{graphicx}
\usepackage[]{caption}
\usepackage{amsmath}
\usepackage{amssymb}
\def\be{\begin{equation}}
\def\ee{\end{equation}}%
\def\nn{\nonumber}
\def\pl{\partial}

\def\bea{\begin{eqnarray}}
\def\eea{\end{eqnarray}}

\def\Riem2{R_{\mu \nu \lambda \rho}^{\hspace{18pt} 2}}
\def\Ricci2{R_{\mu \nu}^{\hspace{9pt} 2}}
\def\F2{F^a_{\mu \nu} F_a^{\mu \nu}}

\begin{document}
\bibliographystyle{prsty}
\title{A Back-Reaction Induced Lower Bound on the Tensor-to-Scalar Ratio}
\author{Patrick Martineau and Robert Brandenberger}
\affiliation{Department of Physics, McGill University, 
3600 University Street,\\ 
Montr\'eal, QC, Canada, H3A 2T8 }
 
\date{\today}
\begin{abstract}
There are large classes of inflationary models, particularly popular
in the context of string theory and brane world approaches to inflation,
in which the ratio of linearized tensor to scalar metric fluctuations
is very small. In such models, however, gravitational waves produced
by scalar modes cannot be neglected. We derive the lower bound on the 
tensor-to-scalar ratio by considering the back-reaction of the scalar 
perturbations as a source of gravitational waves. These results show that 
no cosmological model that is compatible with a metric scalar amplitude of 
$\approx 10^{-5}$ can have a ratio of the tensor to scalar power
spectra less than $\approx 10^{-8}$ at recombination and that higher-order terms leads to logarithmic growth for r during radiation domination. Our lower bound also applies to
non-inflationary models which produce an almost scale-invariant spectrum
of coherent super-Hubble scale metric fluctuations. 

\end{abstract}

\maketitle
\section{Introduction}

Direct measurement of primordial gravitational waves or of the tensor to 
scalar ratio, r, would provide cosmologists with invaluable information 
about the state of the inflationary universe - specifically, the value of 
the Hubble rate. The reason for this is that, unlike detecting the scalar 
amplitude, detecting the tensor amplitude unambiguously determines the value 
of the inflationary scale since \cite{Yao:2006px}
\be
r \, \approx \, \frac{m^{2}_{Pl}}{\pi}(\frac{V'}{V})^{2},
\ee
where slow-roll conditions have been assumed. The ratio r is defined
as the ratio of the power spectra of tensor to scalar modes.

The current experimental upper bound is $r < 0.5$ and it is 
anticipated that, in the next few years, we will be able to probe for values 
of r in the range $10^{-1}-10^{-2}$ \cite{Bock:2006yf}. In the
approximation in which both the scalar and tensor spectra are computed
in linear cosmological perturbation theory, the value of r can be
used to distinguish different classes of inflationary models. Whereas
single scalar field models of large field inflation \cite{Linde83}
typically predict a value of r of around $10^{-2}$ 
(plus or minus an order of magnitude), small field models of
inflation \cite{Linde82,AS} with potentials of Coleman-Weinberg
type \cite{CW} and hybrid inflation models \cite{hybrid}
typically predict a very small amplitude. For example, for a 
potential which in the slow-roll region can be approximated by
\be
V \, = \, V_0 - \lambda \phi^4 \, ,
\ee 
where $\lambda \ll 1$ is a self-coupling constant, the ratio is
\be
r \, \sim \, \lambda^2 ({H \over {m_{Pl}}})^2 \, .
\ee
In the case of hybrid inflation, the small value of r can be seen
either by direct computation, or else by making use of the ``Lyth
bound'' \cite{Lyth}
\be
r \, < \, {8 \over {N^2}} ({{\Delta \phi} \over {m_{Pl}}})^2 \, ,
\ee
where $N$ is the number of e-foldings of inflation between the time
when cosmological scales exist the Hubble radius during inflation and
the end of the inflationary phase, and $\Delta \phi$ is the change
in the value of $\phi$ during the corresponding period of inflation
(which in the case of hybrid inflation is much smaller than $m_{Pl}$.
In particular, since the power spectrum of gravitational waves is
given by the Hubble constant during inflation, low scale models of
inflation can have tiny values of r, values as low as $10^{-24}$
appearing in the literature \cite{Kallosh:2007wm}. Most of the
recently popular string-motivated brane-antibrane inflation models
(see \cite{braneinflation} for recent reviews) are of hybrid inflation
type and hence lead to very small values of r, and in this
context an upper bound on the primordial tensor to scalar ratio
has been proposed \cite{Liam}.

The purpose of this short note is to point out that there exists an 
alternative mechanism for the production of tensors that is completely 
independent of the specific inflationary model. 
The back-reaction of scalar perturbations 
can act as a source of gravitational waves which, despite not being typically 
thought of as a main contributor to r in the case of large field inflationary 
models, could dominate the value of r in other models. The main
contribution to the induced gravitational waves is produced after
reheating. Hence, our analysis and the resulting lower bound on r
applies to all models which produce an almost scale-invariant spectrum
of coherent super-Hubble scale fluctuations.

Note that there has been a substantial body of previous work on
gravitational waves induced at higher order in perturbation theory
\cite{epsilon},\cite{other}, in particular on gravitational waves produced
at the end of inflation \cite{Easther,Uzan}. There has also
been recent work on gravitational waves generated by the curvaton
\cite{curvaton}. What is new in our work
is the focus on the modifications to the tensor to scalar ratio.
Note that the conclusion that secondary gravitational waves may
dominate over the primordial ones has also recently been reached
in \cite{baumann}.

\section{Gravitational Waves from Scalars}

Our starting point is the following perturbed line element written
in terms of conformal time $\eta$, and denoting the scale factor of
the background by $a(\eta)$:
\be
ds^{2} \, = \, a(\eta)^{2}
[(1+2\Phi)d\eta^{2}-(1-2\Phi)\delta_{ij}dx^{i}dx^{j}+h_{ij}dx^{i}dx^{j}].
\ee
The scalar metric fluctuations are described by $\Phi$, a function
of space and time, and the gravitational waves are given by the transverse
and traceless tensor $h_{ij}$. We are using longitudinal gauge to 
describe the linearized scalar metric perturbations
\footnote{At quadratic order, corrections in the
scalar metric fluctuations are induced. In particular, the coefficient
of the spatial diagonal metric perturbation can no longer be identified
with the perturbation of the time-time component (see e.g.
\cite{hwang} for an analysis of cosmological fluctuations and
gravitational wave production at
second order). The second order scalar metric fluctuations will
also induce second order gravitational waves, an effect which
we are neglecting.}, and are
assuming the absence of anisotropic stress (see \cite{MFB} for
a comprehensive review of the theory of cosmological perturbations,
and \cite{RHBrev} for an introductory overview). For each wavenumber $k$,
there are two polarization states for gravitational waves, each
described by an amplitude $h$ and a polarization tensor $\epsilon_{ij}$.

In other words, each second rank tensor $h_{ij}$ can be decomposed as
\be
h_{ij}(\eta,x)\,=\,h_{(1)}(\eta,x)\epsilon^{(1)}_{ij}+h_{(2)}(\eta,x)\epsilon^{(2)}_{ij} \, .
\ee
In what follows, we make use of this decomposition to express the dynamics of the gravitational waves, $h_{ij}$, in terms of the scalar coefficient functions $h^{(1,2)}$.
 
At linear order in the fluctuations, the scalar and tensor modes are
independent. However, if the magnitude of the linearized gravitational
waves is very small compared to the magnitude of the scalar modes,
then the secondary tensor fluctuations generated by non-linear
interactions from the scalar modes cannot be neglected. 
We can determine the coupling between $\Phi$ and $h_{ij}$ by expanding the 
gravitational action in powers of the perturbation amplitude. At leading 
order, we find:
\bea
S_{grav}\,&=&\,\frac{m^{2}_{Pl}}{16\pi}\int{}d^{4}x \sqrt{-g}R\,\label{grav}  
\\
&=&\frac{m^{2}_{Pl}}{16\pi}\int{}d^{4}x\,[\frac{a(\eta)^{2}}{2}((h')^{2}-(\pl_{i}h)^{2})\nn \\
&& - a^{2}(\eta)h(8\Phi\nabla^{2} \Phi+6\,(\pl_{i}\Phi)^{2})+...]
\, .\nn
\eea
In the above (and below), we have supressed the indices denoting the graviton polarization state as both states couple identically to matter.

Additional sources for gravitons arise from the matter sector coupled 
through the kinetic term. In the spirit of simple inflationary universe
models, we will take the matter sector to be described by a single
canonically normalized scalar field $\phi$. In this case we write
\bea
S_{matter}\,\,&=&\,
\int{}d^{4}x \frac{1}{2}\sqrt{-g}(g^{\mu\nu}\pl_{\mu}\varphi\pl_{\nu}\varphi-V(\varphi)]
\label{matter}\\
&=&\int{}d^{4}x \, ...\, 
-\frac{1}{2}\,a^{2}(\eta)h[(\pl_{i}\delta\varphi)^{2}\nn\\
&& + 
2\Phi\pl^{i}\varphi\pl_{i}\delta\varphi+2\Phi^{2}(\pl_{i}\varphi)^{2})
\, ... \, ,\nn
\eea
where we have extracted only the terms which are linear in h.
If the background is homogeneous, then the terms proportional to 
$\pl_{i}\varphi$ are absent.

Combining (\ref{grav}) and (\ref{matter}) leads to the following, 
sourced equation of motion for the tensors:
\bea 
\mu''&+&(-\nabla^{2}-\frac{a''}{a})\mu\,\label{dynamics} \\
&=&\,
\frac{1}{3}\,a(\eta)[16\Phi\nabla^{2} \Phi+12\,(\pl_{i}\Phi)^{2}\nn\\
&& + \frac{16\pi}{m^{2}_{Pl}}(\pl_{i}\delta\varphi)^{2}],\nn
\eea
where we have rescaled the tensor mode, $h=\mu/a(\eta)$. This 
transformation has the advantage of eliminating the Hubble damping term 
and making the particle production process more transparent.

In the standard case, squeezing of the quantum vacuum state leads to 
particle production \cite{Grishchuk:1990bj} with the squeezing being 
determined by the dynamics of the background. The squeezing plays the 
role of a tachyonic instability, which can be seen by the presence of a 
negative mass term in (\ref{dynamics}). Exactly the same thing happens in 
the scalar sector except that, instead of the squeezing term being 
proportional to $-\frac{a''}{a}$, it is proportional to 
$-\frac{z''}{z}$ with $z\,=\,a\frac{\varphi'}{\mathcal{H}}$. It is this 
different dependence on the behaviour of the inflaton that makes it 
possible to have a scenario in which the tensor amplitude is far below 
that of the scalar.

We now consider a scenario in which the squeezing produces a negligible 
amount of gravitons, but a large amplitude of scalar metric fluctuations.
In this case, we can determine the amplitude of the (secondary) tensors 
being produced via interactions from the scalar modes (in the language
of particle physics, one could say ``by scalar decay''). To this end, we 
neglect the squeezing term in (\ref{dynamics}) entirely. Since we
focus on super-Hubble scale fluctuations, we can also neglect the
spatial gradient terms \footnote{A formalism which does not make
these two approximations and thus applies also to short wavelength
fluctuations was recently discussed in \cite{Uzan}. The full
partial differential equation can be solved by a Greens function
method. The solution involves an integral over momenta and time.
Applied to oscillatory fluctuations, the time integral produces
a delta function which leads to the conclusion that short wavelength
scalar modes do not contribute to the production of gravitational
waves. For long wavelength fluctuations, this delta function does
not appear, and, as we show below, gravitational waves are indeed
generated. One way of convincing oneself that this must
be possible is that the coupling we are analyzing is that between
two vectors (the gradients of the scalar field) and a tensor,
and this coupling in general does not vanish. However, the analysis 
of \cite{Uzan} gives us a
further argument to focus exclusively on super-Hubble modes.}.
With these approximations, the partial differential equation
(\ref{dynamics}) reduces to an ordinary one which takes the form
\bea
\mu''\,\label{dynamics2} 
&=&\,
\frac{1}{3}\,a(\eta)[16\Phi\nabla^{2} \Phi+12\,(\pl_{i}\Phi)^{2}\nn\\
&& + \frac{16\pi}{m^{2}_{Pl}}(\pl_{i}\delta\varphi)^{2}].
\eea

The above equation (\ref{dynamics2}) makes it clear that scalar metric
fluctuations source gravitational waves at all times.
We can simplify the equation by noting that the scalar sector has 
only one independent degree of freedom \cite{MFB}. During
slow-roll inflation we can make use of the 
constraint equation \cite{Abramo:1997hu}
\be
\delta\varphi \, \simeq \, -2\frac{V}{V'}\Phi,
\ee
to recast (\ref{dynamics2}) as
\bea
\mu'' &\simeq& \label{dynamics3} 
\frac{4}{3}a(\eta)[4\Phi\nabla^{2}\Phi\\
&& + (3+\frac{16\pi}{m^{2}_{Pl}}(\frac{V}{V'})^{2}(\pl_{i}\Phi)^{2}].\nn
\eea
After reheating, a formula analogous to (\ref{dynamics2}) but with
matter taken to be a perfect fluid must be used.

We first focus on the gravitational waves generated during inflation.
Making use of the definition of the slow-roll parameter
\be \label{eps}
\epsilon \, \equiv \, \frac{m^{2}_{Pl}}{16\pi}(\frac{V'}{V})^{2}
\ee
in (\ref{dynamics3}), we see something perhaps a little surprising: the second 
term in the source (the term originating from the matter sector) 
scales as $1/\epsilon$. As mentioned above, in the standard case 
(ignoring back-reaction), a small value of $\epsilon$ suppresses the 
amplitude of the tensors. However, including the effects of back-reaction, 
we see that a small $\epsilon$ leads to a higher amplitude of
induced gravitational waves. 

We now wish to use the above formalism to compute the amplitude of
the secondary gravitational waves at the time when the mode of interest
is crossing the Hubble radius during the initial inflationary phase, assuming
that the initial state of all fluctuation modes on ultraviolet scales
is the quantum vacuum state. Because of our assumption on the initial state,
it is only scalar modes which have already exited the Hubble radius and
thus undergone some squeezing which can produce such secondary
gravitational waves.

One way to compute the source term in (\ref{dynamics3}) is to expand
$\Phi$ in Fourier modes, insert into the expression on the right hand side
of the equation, use the fact that the scalar spectrum is
scale-invariant on super-Hubble scales, and 
to extract the $k$'th Fourier mode of $\mu$ by inverse
Fourier transform of the result and by contracting with the polarization 
tensor of the gravitational wave. The integrals over momenta run from
an infrared cutoff (which can be taken to be the scale corresponding to
the Hubble radius at the beginning of inflation, but which does not play
any role in this calculation) to momenta corresponding to the Hubble
radius. Modes on sub-Hubble scales are in their vacuum state and hence
cannot generate any gravitational waves. Also, as shown in \cite{Uzan},
the effect of short wavelength scalar fluctuations would average to
zero in our coupling term.

A short cut to the calculation is to simply to
promote $\Phi$ to the status of quantum operator \cite{MFB} and to 
take the vacuum expectation value of the right hand side of (\ref{dynamics3}),
making use of the two-point function (in the limit that $r \to 0$)
\be
\langle 0|\Phi(x,\eta)\Phi(x+r,\eta)|0 \rangle \, = \,
\int_{0}^{\infty}\,\frac{dk}{k}\frac{\sin(kr)}{kr}|\delta_{\Phi}|^{2},
\ee
and restricting the integral to the IR phase space, $k=[0,\mathcal{H}]$,
where $\mathcal{H}$ is the Hubble rate in conformal time. In the above,
$|\delta_{\Phi}|^{2}$ is the power spectrum of the scalar modes (including
the phase space factor $k^3$). Multiplying with the polarization tensor
to extract the amplitude $\mu$ of the gravitational wave, we thus get
\be \label{dynamics4}
\mu''\, \simeq \, 
-\frac{2}{3}a(\eta)\,[7+\frac{1}{\epsilon}]\mathcal{H}^{2}|\delta_{\Phi}|^{2}
\, ,
\ee
through which the evolution of $\mu$ is now directly related to the 
scalar power spectrum $|\delta_{\Phi}|^2$. Note that in the above
formula, the scalar metric power spectrum is the one during inflation. 

Several remarks need to be made about this calculation: to obtain this result, 
we assumed scale-invariance of the scalars. As well, (\ref{dynamics4}) 
can only be used to extract information about $\mu$ on large scales, i.e. 
$H \gg k_{phys}$. The physical reason for this condition is obvious -
momentum conservation forces the produced tensors to have smaller momenta 
than the scalars from which they originate. Considering contributions 
arising solely from the scalar IR sector is not as limiting as it may seem 
due to the fact, through the course of inflation, the IR phase space 
grows exponentially while the UV phase space remains constant 
(see the arguments presented in \cite{Abramo:1997hu} and 
\cite{Mukhanov:1996ak}). However, in our calculation there is no IR
divergence and the non-gradient contributions of the IR modes (those
which in the case of adiabatic matter fluctuations are pure gauge
\cite{Unruh,Ghazal,Abramo}) do not contribute. Note that
the $\frac{1}{\epsilon}$ dependence on the size of the effects of 
back-reaction has also been noticed in \cite{epsilon}, \cite{Losic}, and in a very different context,\cite{Martineau:2006ki}.

Integrating (\ref{dynamics4}) immediately yields
\be
\mu \, \simeq \,
-\frac{a(\eta)}{3}[7+\frac{1}{\epsilon}]|\delta_{\Phi}|^{2},
\ee
or
\be \label{hfinal}
h \, \simeq \, -\frac{1}{3}[7+\frac{1}{\epsilon}]|\delta_{\Phi}|^{2},
\ee
where we have used $a(\eta)\,=\,\frac{-1}{H \eta }$, and 
$\mathcal{H}\,=\,a(\eta)H$. The scale invariance of the above is manifest.

The amplitude of the power spectrum of scalar metric perturbations 
during inflation is suppressed compared to the spectrum after
inflation by the factor $\epsilon^2$, the inverse of the squeezing factor
of $\Phi^2$ during the transition between the inflationary phase
and the radiation phase of Standard Cosmology. Taking the current
power spectrum to be $\approx 10^{-10}$, we estimate the contribution 
to the tensor-scalar ratio induced by back-reaction before reheating to be
\be
r \, \approx \, 10^{-10} \epsilon^2 \, .
\label{final}
\ee
This is negligible compared to the contribution to $r$ of the linear
gravitational waves produced from quantum vacuum fluctuations.

Next, we consider the gravitational waves induced by back-reaction
after reheating. To obtain a lower bound on the effects of back-reaction,
we neglect the matter contribution on the right hand side of 
(\ref{dynamics2}). Hence, there is no $\epsilon^{-1}$ enhancement
factor. However, the amplitude of $\Phi$ to be used is not suppressed
by $\epsilon$. Hence, repeating the steps which led to (\ref{hfinal})
leads to an analogous formula without the $\epsilon^{-1}$ term in the
square parentheses on the right-hand side of the equation. This
give a contribution
\be \label{final2}
r \, \approx \, 10^{-10} \,  .
\ee
Hence, the back-reaction of scalar metric fluctuations has a more
important effect after reheating than before.

Our analysis is not restricted to the inflationary era. Interestingly, (\ref{dynamics4}), predicts the creation of gravitons on super-Hubble scales even after reheating. Direct substitution of the radiation-domination scale factor ($a(\eta)\,=\,\eta/\eta_{0}$) leads to the result 

\be{h(\eta)\, \sim\, \ln(a(\eta)).}\ee

At the onset of matter domination ($a(\eta)\,\sim\,\eta^{2}$), the amplitude of the super-Hubble tensor modes again becomes constant.

Taking the above into consideration, we take as our final estimate for the back-reaction determined tensor-to-scalar ratio at the time of recombination to be

\be{r\, \approx \, 10^{-8}.}\ee

At the order in perturbation theory we are now working in we must
also consider the decay of the gravitational waves into 
scalars through interaction terms of the form
\be
S_{h^{2} \to \Phi} \, \propto \,
\int{}d^{4}x a^{2}(\eta)\Phi h_{j}^{i}\nabla^{2}h_{i}^{j} \, . 
\ee 
However, for the values of $r$ obtained here, this decay is
negligible.

\section{Conclusion}

We have estimated the amplitude of gravitational waves produced via
the back-reaction of scalar metric fluctuations, focusing on their
effect on the tensor-to-scalar ratio $r$.
In conclusion, our result (\ref{final2}) shows that back-reaction of scalar
modes will produce a sizeable gravitational wave background in a wide variety 
of inflationary models in which the value of $r$ computed from purely
linear theory is very small. 

The contribution to $r$ from back-reaction of scalar modes picks up
its major contribution after reheating. Hence, our lower bound on $r$
will be valid in any theory producing a roughly scale-invariant
spectrum of fluctuations at late times, provided these fluctuations
are coherent outside the Hubble radius (which appears to be demanded
by the data based on the acoustic peak structure of the observed
angular power spectrum of cosmic microwave anisotropies). Examples
of such theories are the Ekpyrotic universe scenario \cite{Ekp}, the
Pre-Big-Bang model \cite{PBB} and the recently proposed string gas
structure formation paradigm \cite{NBV}. Our analysis is
particularly relevant for the Ekpyrotic scenario in which a negligible
amount of gravitational waves are generated in the linear theory.
Back-reaction, however, will induce a contribution which is similar in
magnitude to what is induced in many small field inflationary models.

An interesting point that was noted was the ability of backreaction to increase the amplitude of IR modes after inflation. This is in stark contrast with the linear result in which the amplitudes of both scalar and tensor modes are frozen until the time of second Hubble crossing.

This work is complementary to an earlier study of the back reaction effects
of scalar modes onto the scalar modes themselves \cite{Patrick} which were
shown to be negligible. As this note was being finalized, a paper
appeared \cite{Clarkson} which studies vector modes produced by 
primordial density fluctuations.

\begin{acknowledgments}

The authors would like to thank Aaron Berndsen for useful discussion.
We are grateful to Cristian Armendariz-Picon for pointing out
an error in an earlier version of this manuscript.
This work was supported by an NSERC Discovery Grant,
by the Canada Research Chairs program and
by funds from a FQRNT Team Grant. 

\end{acknowledgments}

\end{document}